\documentclass[a4paper,11pt]{article}
\pdfoutput=1

\usepackage{jcappub}
\usepackage{subcaption}
\usepackage{units}
\usepackage{mathrsfs}
\usepackage{mathtools}
\usepackage{hyperref}
\usepackage{gensymb}
\usepackage{bm}
\usepackage{tabularx,caption,booktabs}
\usepackage{comment}
\usepackage[dvipsnames]{xcolor}
\usepackage[normalem]{ulem}
\usepackage[T1]{fontenc}
\usepackage{lineno}

\title{\boldmath Combined search in dwarf spheroidal galaxies for branon dark matter annihilation signatures with the MAGIC telescopes}

\author [1] {S.~Abe}
\author [2] {J.~Abhir}
\author [3] {A.~Abhishek}
\author [4] {V.~A.~Acciari}
\author [5] {A.~Aguasca-Cabot}
\author [6] {I.~Agudo}
\author [7] {T.~Aniello}
\author [8,40] {S.~Ansoldi}
\author [9] {A.~Arbet Engels}
\author [10] {C.~Arcaro}
% ...
\author{(the MAGIC Collaboration)}
\author[49,50,51,*]{V. Gammaldi}
\author[14]{D. Nieto}

\affiliation [1] {Japanese MAGIC Group: Institute for Cosmic Ray Research (ICRR), The University of Tokyo, Kashiwa, 277-8582 Chiba, Japan}
\affiliation [2] {ETH Z\"urich, CH-8093 Z\"urich, Switzerland}
\affiliation [3] {Universit\`a di Siena and INFN Pisa, I-53100 Siena, Italy}
\affiliation [4] {Institut de F\'isica d'Altes Energies (IFAE), The Barcelona Institute of Science and Technology (BIST), E-08193 Bellaterra (Barcelona), Spain}
\affiliation [5] {Universitat de Barcelona, ICCUB, IEEC-UB, E-08028 Barcelona, Spain}
\affiliation [6] {Instituto de Astrof\'isica de Andaluc\'ia-CSIC, Glorieta de la Astronom\'ia s/n, 18008, Granada, Spain}
\affiliation [7] {National Institute for Astrophysics (INAF), I-00136 Rome, Italy}
\affiliation [8] {Universit\`a di Udine and INFN Trieste, I-33100 Udine, Italy}
\affiliation [9] {Max-Planck-Institut f\"ur Physik, D-85748 Garching, Germany}
\affiliation [10] {Universit\`a di Padova and INFN, I-35131 Padova, Italy}
% ...
\affiliation [14] {IPARCOS Institute and EMFTEL Department, Universidad Complutense de Madrid, E-28040 Madrid, Spain}
% ...
\affiliation [40] {also at International Center for Relativistic Astrophysics (ICRA), Rome, Italy}
% ...
\affiliation[49]{Departamento de F\'{i}sica Te\'{o}rica, Facultad de Ciencias, Mod. 15,  Universidad Autónoma de Madrid, E-28049 Madrid, Spain}
\affiliation[50]{Instituto de F\'{i}sica Te\'{o}rica, UAM-CSIC, Calle Nicol\'{a}s Cabrera 13-15, Campus de Cantoblanco, E-28049 Madrid, Spain}
\affiliation[51]{Now at Department of Information Technology, Escuela Politecnica Superior, Universidad San Pablo-CEU, CEU Universities, Campus Monteprincipe, Boadilla del Monte, Madrid 28668, Spain}

\affiliation[*]{Corresponding authors: T. Miener, D. Kerszberg, V. Gammaldi.}

\emailAdd{contact.magic@mpp.mpg.de}
\emailAdd{viviana.gammaldi@ceu.es}

\abstract{Massive brane fluctuations, called branons, behave as weakly interacting massive particles, which is one of the most favored class of candidates to fulfill the role of the dark matter (DM), an elusive kind of matter beyond the Standard Model. We present a multi-target search in dwarf spheroidal galaxies for branon DM annihilation signatures with a total exposure of 354 hours with the ground-based gamma-ray telescope system MAGIC. This search led to the most constraining limits on branon DM in the sub-TeV and multi-TeV DM mass range. Our most stringent limit on the thermally-averaged annihilation cross-section (at $95\%$ confidence level) corresponds to $ \langle \sigma v \rangle \simeq \unit[1.9 \times 10^{-24}]{cm^{3}s^{-1}} $ at a branon mass of $ \sim \unit[1.5]{TeV}$.}

\begin{document}
\maketitle
\flushbottom

\section{Introduction}

\label{sec:intro}

The nature of dark matter (DM) is still an open question for modern physics. This non-baryonic and non-relativistic kind of matter is suggested to be accountable for $84\%$ of the matter density of the Universe~\cite{Aghanim:2018eyx}. Among many other DM candidates~\cite{Bertone:2005xv}, massive brane fluctuations (branons) emerging from the brane-world theory~\cite{2003PhRvL..90x1301C} have been proposed as DM candidates, since their characteristics match the ones of weakly interacting massive particles (WIMPs)~\cite{2012PhRvD..86b3506S}. Gamma-ray telescopes could potentially detect DM, in particular branons, indirectly by observing photons, e.g. via quark hadronization or final state radiation from charged particles, in the very-high energy (VHE, $ \gtrsim \unit[50]{GeV} $) domain of astrophysical regions presenting large DM densities.

Dwarf spheroidal galaxies (dSphs) are preferred targets for indirect DM searches. Contrary to the Galactic Center (GC) and galaxy clusters, also very prominent targets for DM searches~\cite{2002PhRvL..88s1301M,2007ApJ...654..897B}, dSphs are not expected to host any strong conventional gamma-ray emitter that may hinder the detection of a subdominant DM signal. In addition, dSphs are not as extended targets as the GC or galaxy clusters, with source angular extension presenting a challenge for ground-based gamma-ray telescopes due to their angular resolution. Nevertheless, the determination of the DM content in those objects remains the most uncertain ingredient in most DM analyses. Previous work, the first branon DM search in the VHE domain with the Major Atmospheric Gamma Imaging Cherenkov (MAGIC) telescopes~\cite{Miener:2022fon,2022arXiv220507055M}, utilized the deepest exposure on any single dSph to date, namely Segue~1. Being an interesting target for DM searches, the DM content of Segue~1 has been an active subject of debate~\cite{2016MNRAS.462..223B}. A refined analysis was carried out in~\cite{Bonnivard:2015xpq,2016MNRAS.462..223B} with a more accurate determination of member stars in Segue~1, leading to a lower and more uncertain value for its DM content than previously determined in~\cite{2015ApJ...801...74G}. With the aim of providing more robust and more constraining results, in this work we performed a multi-target branon DM search using two independent determinations of the DM content from the literature.

The present article provides a search for branon DM with gamma rays by combining MAGIC observations of dSphs with a total accumulation of 354 hours leading to the most stringent and robust branon DM limits in the sub-TeV and multi-TeV DM mass range to date. In Sec.~\ref{sec:BranonDM}, we briefly review the brane-world theory and we present the expected photon flux from branon DM annihilation. The observational campaigns on dSphs by the MAGIC telescopes and the adopted estimations of the DM content in those objects are summarized in Sec.~\ref{sec:Observations}. The analysis methodology is explained in Sec.~\ref{sec:Analysis}, while the final results in terms of upper limits (ULs) on the annihilation cross section of branon DM particles and the tension of the brane are presented in Sec.~\ref{sec:Results}. Finally, we discuss the ULs and compare them with the model-independent ULs from MAGIC~\cite{MAGIC:2021mog} in Sec.~\ref{sec:DisCon}.

\section{Gamma rays from branon annihilation}

\label{sec:BranonDM}

Standard Model (SM) fields could exist on a tridimensional brane embedded in a higher dimensional spacetime, where gravity propagates. These extra-dimensional models~\cite{1998PhLB..429..263A,1999PhRvD..59h6004A} were originally proposed as a potential solution to the hierarchy problem. However, they also provide us with natural DM particle candidates. In the context of the \textit{brane-world scenario} with low~\footnote{The low brane tension regime refers to when the tension of the branon is greater than the mass of the branon, where the branon dynamic is given by the Nambu-Goto action added to the usual SM action~\cite{2020PDU....2700448C}.} brane tension $f$, massive brane fluctuations (branons) in the direction of the N extra-dimensions are natural DM candidates~\cite{2003PhRvL..90x1301C,2003PhRvD..68j3505C}. The lowest-order effective Lagrangian for branon DM reads~\cite{2003PhRvD..67g5010A, 2012PhRvD..85d3505C}

%Roughly speaking, brane-world scenarios address the hierarchy problem in Physics by suggesting that our Universe is a (3 + 1)-dimensional brane, where the SM fields propagate. Such a brane is embedded in a higher extra-dimensional compact space - the bulk - that hosts the propagation of gravity. When a curvature is produced along the extra-dimensions the isometry of the brane - that is considered as gauge symmetry -  is broken and the brane fluctuation can be parametrised by a massive (M) pseudo-Goldstone field, πα, dubbed branon. If the tension of the branon is bigger than the mass of the branon, we are in the low brane tension regime, where the branon dynamic is given by the Nambu-Goto action added to the usual SM action. When f< M we are in the strongly coupled region, i.e. our exclusion limits could change due to the validity of the Effective Field Theory.\cite{2020PDU....2700448C}

\begin{equation}
\label{eq:lagrangian}
    \mathscr{L} = \frac{1}{2} g^{\mu\nu} \partial_{\mu} \pi^{\alpha} \partial_{\nu} \pi^{\alpha} - \frac{1}{2} m_{\chi}^{2}\pi^{\alpha}\pi^{\alpha} + \frac{1}{8f^{4}} \left( 4 \partial_{\mu} \pi^{\alpha} \partial_{\nu} \pi^{\alpha} - m_{\chi}^{2}\pi^{\alpha}\pi^{\alpha} g_{\mu\nu} \right) T_{\text{SM}}^{\mu\nu},
\end{equation}
\noindent
where $\pi$ is the branon field and ${\alpha}$ runs over the $ N $ extra-dimensions, $m_{\chi}$ is the particle mass of the branon, and $T^{\mu\nu}_{\text{SM}}$ is the energy-momentum tensor of the SM fields. The coupling of the branons to the SM particles is suppressed by the fourth power of the tension of the brane, rendering them as WIMPs.

The expected differential photon flux produced by branon DM annihilation~\cite{2012PhRvD..85d3505C} is composed of the two terms: (i) the \textit{astrophysical} factor (\textit{J}-factor; see Sec.~\ref{sec:Observations}), which depends on both the distance $ l $ to the target, and the DM distribution $ \rho_{\text{DM}} $ at the source region denoted by its subtended solid angle $\Delta\Omega$, and (ii) the particle physics factor, which includes the branon DM annihilation photon yield. It reads as
\begin{equation}
    \label{eq:Branon_Flux}
    \frac{\text{d}\Phi}{\text{d}E}\left( \langle\sigma v\rangle \right) = \frac{1}{4\pi} \underbrace{ \int_{\Delta\Omega} d\Omega' \int_{\text{l.o.s.}} dl \, \rho_{\text{DM}}^{2} (l,\Omega')}_{\text{\textit{J}-factor}} \cdot \underbrace{ \frac{\langle\sigma v\rangle}{2m^{2}_{\chi}} \sum_{i=1}^{n} \text{Br}_{i} \frac{\text{d}N_{i}}{\text{d}E}}_{\text{DM annihilation}},
\end{equation}
\noindent
where $ \langle\sigma v\rangle $ is the thermally-averaged annihilation cross section and $ \text{l.o.s.} $ stands for line-of-sight. The branon branching ratios $ \text{Br}_{i} $ as a function of $ m_{\chi} $ have been calculated following the prescriptions in~\cite{2012PhRvD..85d3505C} including annihilation into the SM pairs $ W^{+}W^{-} $, $ ZZ $, $ hh $, $ e^{+}e^{-} $, $ t\bar{t} $, $ c\bar{c} $, $ \mu^{+}\mu^{-} $, $ \tau^{+}\tau^{-} $ and $ b\bar{b} $. However, the $ W^{+}W^{-} $, $ ZZ $ and $ hh $ channels are the dominant contributors in our search for TeV branon DM. The differential photon yields per annihilation $ \text{d}N_{i}/\text{d}E $, including electroweak (EW) corrections, are taken from the PPPC 4 DM ID distribution for this work~\cite{2011JCAP...03..051C}. The resulting differential photon yield per branon annihilation for a set of branon DM masses can be found in ~\cite{2020JCAP...10..041A,Miener:2022fon,2022arXiv220507055M,Miener:2022zws}.\newpage

\section{dSph observations by the MAGIC telescopes}

\label{sec:Observations}

The \textit{Florian Goebel} Major Atmospheric Gamma-ray Imaging Cherenkov (MAGIC) telescopes\footnote{\href{https://magic.mpp.mpg.de}{https://magic.mpp.mpg.de}} are two 17-m diameter reflector imaging atmospheric Cherenkov telescopes (IACTs) situated at an altitude of 2200~m~a.s.l. at the Roque de los Muchachos Observatory ($ 28.8^{\circ}$ N, $ 17.9^{\circ}$ W) on the Canary Island of La Palma, Spain. MAGIC inspects the VHE gamma-ray sky (above $\gtrsim\unit[30]{GeV}$) with a $ 3.5\degree $ field of view probing the most extreme astrophysical environments in our universe. The point-source 5$\sigma$ sensitivity above $\unit[220]{GeV}$ of MAGIC is $\sim0.7\%$ of the Crab Nebula flux for 50~h of observations near zenith with an associated energy resolution of $\sim$ 16\% and a $0.07\degree$ angular resolution measured as the 68\% containment radius of the gamma-ray excess. A detailed performance study of the MAGIC telescopes can be found in~\cite{2016APh....72...76A}.

The MAGIC Collaboration has carried out extensive observational campaigns on dSphs in the Northern Hemisphere throughout the years, motivated by the search for DM signals in these objects. At first, MAGIC observed the dSphs Draco, Willman~1, and Segue~1 with the MAGIC-I telescope in single telescope mode around 2009~\cite{2009arXiv0907.0738L,2011JCAP...06..035A}. Those data have not been used in this work because of the superseding sensitivity by the additional MAGIC-II telescope. After upgrading MAGIC to a stereoscopic IACT system, Segue~1 was observed between 2011 and 2013 with an exposure of 158 hours~\cite{2014JCAP...02..008A,2016JCAP...02..039M}. This is still the deepest observation of any dSph by an IACT to date. Additionally, MAGIC observed Ursa~Major~II between 2014 and 2016~\cite{2018JCAP...03..009A}, Draco in 2018, and Coma~Berenices in 2019 leading to a total accumulation of 354 hours of dSph observations~\cite{MAGIC:2021mog}. The observation of Triangulum~II by the MAGIC telescopes~\cite{MAGIC:2020ceg} has been excluded due to the statistical uncertainty on the determination of the DM distribution in this object~\cite{2017ApJ...838...83K}. The dSph observations by the MAGIC telescopes included in this combined search for branon DM are summarized in Tab.~\ref{Tab:MAGICObs}. No effects of the extragalactic background light absorption~\cite{SaldanaLopez:2021} are considered in our analysis, since the observed dSphs are positioned only a few tens of kpc away from us (see Tab.~\ref{Tab:dSphs}).

\begin{table}[h]
\centering
\caption{Summary of the considered dSph observations by the MAGIC telescopes~\cite{MAGIC:2021mog}. We report the zenith distance (zd) range, the total observation time ($ T_{\text{obs}} $), and the energy range (E). We also list the angular radius ($\theta$) of the signal region and the significance of detection (S$_{\text{Li\&Ma}}$) calculated by following Li\&Ma~\cite{LiMa:1983}. Please note that the significance of detection is not reported for Coma~Berenices and Draco in~\cite{MAGIC:2021mog}, but no gamma-ray excess has been found.}
%\footnotesize
%the normalization between background and signal regions ($ \tau $),
\begin{tabular}{cccccc}
\hline
\hline
%\CellTopTwo{}
Name & zd [\degree] & $ T_{\text{obs}} $ [h] & E [TeV] & $\theta$ [\degree] & S$_{\text{Li\&Ma}}$[$\sigma$] \\
\hline
%\CellTopTwo{}
Coma~Berenices & $5 - 37 $ & $49$ & $0.06 - 10$ & $0.17$ & $0.8$ \\
%\CellTopTwo{}
Draco & $29 - 45$ & $52$ & $0.07 - 10$ & $0.22$ & $-0.7$ \\
%\CellTopTwo{}
Segue~1 & $13-37$ & $158$ & $0.06 - 10$ & $0.12$ & $-0.5$ \\
%\CellTopTwo{}
Ursa~Major~II & $35-45$ & $95$ & $0.12 - 10$ & $0.30$ &  $-2.1$ \\
\hline
\hline
%\CellTopTwo{}
\end{tabular}
\label{Tab:MAGICObs}
\end{table}

\begin{figure}[h]
    \centering
    \includegraphics[scale=0.5]{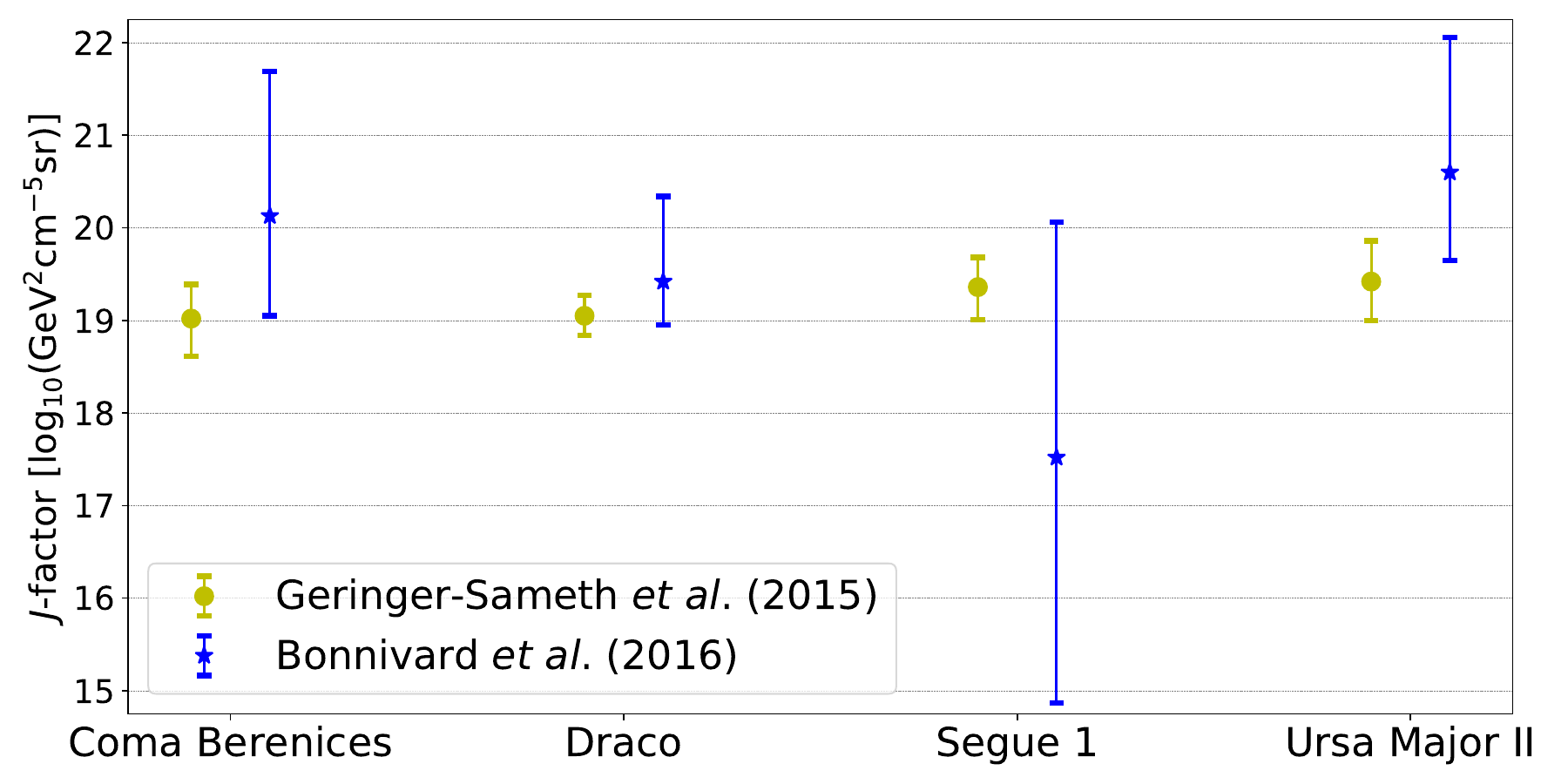}
    \caption{Comparison of the \textit{J}-factor values and its uncertainties from GS15~\cite{2015ApJ...801...74G} (green dots) and B16~\cite{Bonnivard:2015xpq} (blue stars) for all considered dSphs.}
    \label{fig:Jfactors}
\end{figure}

The driving factor of the search for DM annihilation signatures in dSphs is the estimation of the DM content in those objects. This is a challenging task resulting in rather large uncertainties, which are dominant in our analysis. Therefore, this work includes a systematic study on the impact of the estimation of the \textit{J}-factors to our derived constraints by performing the same likelihood analysis (see Sec.~\ref{subsec:LklAnalysis}) with two different sets of the \textit{J}-factors from the literature, i.e. Geringer-Sameth \emph{et al.}~\cite{2015ApJ...801...74G} (henceforth referred to as GS15) and Bonnivard \emph{et al.}~\cite{Bonnivard:2015xpq} (henceforth referred to as B16). The derivation of the two \textit{J}-factor sets was carried out in~\cite{2015ApJ...801...74G,Bonnivard:2015xpq} using a Jeans analysis~\cite{Bonnivard:2014kza} of the same kinematic stellar data for Segue~1, Ursa~Major~II, and Coma~Berenices (a detailed description can be found in~\cite{2015ApJ...801...74G}). B16~\cite{Bonnivard:2015xpq} adopted the kinematic stellar data for the classical dSph Draco from~\cite{2015MNRAS.448.2717W}, which differs from the data used by~\cite{2015ApJ...801...74G}. The main differences between the two approaches are the selection of the DM density, velocity anisotropy, and light profiles, as well as the inclusion of systematic uncertainties in~\cite{Bonnivard:2015xpq}. The \textit{J}-factor values and its uncertainties are listed in Tab.~\ref{Tab:dSphs} and visualized in Fig.~\ref{fig:Jfactors}. The largest discrepancy is found for the \textit{J}-factor of Segue~1, since the Jeans analysis in B16~\cite{Bonnivard:2015xpq} is extremely sensitive on the determination of the member stars. The contamination of the dSph stellar sample by a foreground population with different velocity properties complicates membership determination~\cite{2016MNRAS.462..223B} leading to an artificially inflated \textit{J}-factor estimation in GS15~\cite{2015ApJ...801...74G}, where the foreground population is wrongly included in the Jeans analysis.

\begin{table}[h]
\centering
\caption{Summary of the dSph properties. We report the heliocentric distance and Galactic coordinates of each dSph, as well as the total \textit{J}-factor values and its $\pm 1\sigma$ uncertainties from GS15~\cite{2015ApJ...801...74G} and B16~\cite{Bonnivard:2015xpq} used in the present work. The maximum angular distance $\theta_{\rm{max}}$ is the angular distance from the center to the outermost member star considered in the \textit{J}-factor calculation.}
\begin{tabular}{ccccccc}
\hline
\hline
Name & Distance & $l, b$ & $\log_{10}J (\theta_{\rm{max}})$~\{GS15\} & $\log_{10}J(\theta_{\rm{max}})$~\{B16\} \\
& [kpc] &  [$^{\circ}$] & [$\log_{10}(\rm{GeV}^2 \rm{cm}^{-5}\rm{sr})$] & [$\log_{10}(\rm{GeV}^2 \rm{cm}^{-5}\rm{sr})$] \\
\hline
Coma~Berenices & $44$ & $241.89,\: 83.61$ & $19.02^{+0.37}_{-0.41}$ & $20.13^{+1.56}_{-1.08}$ \\
Draco & $76$ & $86.37,\: 34.72$ & $19.05^{+0.22}_{-0.21}$ &  $19.42^{+0.92}_{-0.47}$ \\
Segue~1 & $23$ & $220.48,\: 50.43$ & $19.36^{+0.32}_{-0.35}$ &  $17.52^{+2.54}_{-2.65}$ \\
Ursa~Major~II & $32$ & $152.46,\: 37.44$ & $19.42^{+0.44}_{-0.42}$ & $20.60^{+1.46}_{-0.95}$ \\
\hline
\hline
\end{tabular}
\label{Tab:dSphs}
\end{table}

\section{Analysis technique}

\label{sec:Analysis}

\subsection{Data reduction}
\label{subsec:DataReduction}
The low-level data reduction of the four dSph observations was performed in~\cite{MAGIC:2021mog} using the standard MAGIC analysis software MARS~\cite{2013ICRC...33.2937Z}. We re-analyse the resulting high-level data of four dSph observations~\cite{MAGIC:2021mog} in the context of brane-world extra-dimensional theories using the same open-source analysis software tools~\cite{2021arXiv211201818M} for multi-instrument and multi-target DM searches \texttt{gLike}\footnote{\href{https://github.com/javierrico/gLike}{https://github.com/javierrico/gLike}}~\cite{javier_rico_2021_4601451} and \texttt{LklCom}\footnote{\href{https://github.com/TjarkMiener/likelihood_combiner}{https://github.com/TjarkMiener/likelihood\_combiner}}~\cite{tjark_miener_2021_4597500}.

\subsection{Likelihood analysis}
\label{subsec:LklAnalysis}

In this work, we used the same likelihood analysis as~\cite{MAGIC:2021mog}, which was originally proposed in~\cite{2012JCAP...10..032A}, further utilized in ~\cite{2016JCAP...02..039M,2018JCAP...03..009A} and discussed in~\cite{Rico:2020vlg}. In order to obtain the expected branon DM signal for the MAGIC telescopes, the theoretical branon DM flux (Eq. \ref{eq:Branon_Flux}) is convolved with the instrument response functions (IRFs) for the signal (ON) region $ \text{IRF}_{\text{ON}} \left( E, E'\right) $ including the morphology of the dSph via the \textit{"Donut" MC method}~\cite{2018JCAP...03..009A}, which can be described by the PDF for the energy estimator and the effective collection area. The \textit{Donut MC method} is the procedure to build the IRFs with a specific MC sample representing the source morphology rather than using MC for an assumed point-like source. The specific MC samples are produced by selecting events from the diffuse MC resulting in a donut-shaped distribution. $ E $ and $ E' $ are the true and estimated energy of the gamma-ray photon, respectively. The expected number of signal events in the $ i $-th energy bin yields
\begin{equation}
    \label{expected_numbers_of_signal_events}
    s_{i}(\langle \sigma v \rangle) = T_{\text{obs}} \int_{E_{\text{min},i}}^{E_{\text{max},i}} \text{d}E' \int_{0}^{\infty} \frac{\text{d}\Phi(\langle \sigma v \rangle)}{\text{d}E} \text{IRF}_{\text{ON}} \left( E, E'\right) \text{d}E,
\end{equation}
\noindent
where $ T_{\text{obs}} $ is the total observation time and $ E_{\text{min},i} $ and $ E_{\text{max},i} $ are the lower and upper limits of the $ i $-th energy bin. The thermally-averaged annihilation cross section $ \langle\sigma v\rangle $ is our parameter of interest and therefore the only free parameter in our likelihood analysis.

The binned ($ N_{\text{bins}} = 30 $) likelihood function of the dataset $ \bm{\mathcal{D}'} $ with nuisance parameters $ \bm{\nu} $ reads as:
\begin{equation}
    \label{eq:Binned_lkl}
    \begin{split}
    \mathcal{L}_{\text{bin}} \left( \langle \sigma v \rangle ; \bm{\nu} \mid \bm{\mathcal{D}'} \right) = \prod_{i=1}^{N_{\text{bins}}} \Big[ \mathcal{P} (s_{i}(\langle \sigma v \rangle) + b_{i} \mid N_{\text{ON},i}) \cdot \mathcal{P} (\tau b_{i} \mid N_{\text{OFF},i}) \Big] \times \mathcal{T} \left( \tau \mid \tau_{\text{o}}, \sigma_{\tau} \right)
    \end{split}
\end{equation}
\noindent
where $ \mathcal{P} (x | N) $ stands for a Poisson distribution with mean $ x $ and measured value $ N $, while $ N_{\text{ON},i} $, $ N_{\text{OFF},i} $ are the total number of observed events in the $ i $-th energy bin in the signal (ON) and background (OFF) regions, respectively. The background events in the 30 bin in energy and the normalization between background and signal regions $ \tau $ are nuisance parameters, which leads to a total of 31 nuisance parameters in Eq.~\ref{eq:Binned_lkl}. The likelihood function $ \mathcal{T} \left( \tau \mid \tau_{\text{o}}, \sigma_{\tau} \right) $ is a Gaussian with mean $ \tau_{\text{o}} = 1.0 $ and variance $ \sigma_{\tau}^{2} $, which include statistical and systematic uncertainties on $\tau$ following $\sigma_{\tau} = \sqrt{\sigma_{\tau_{\mathrm{stat}}}^{2} + \sigma_{\tau_{\mathrm{syst}}}^{2}}$. Based on a dedicated performance study of the MAGIC telescopes~\cite{2016APh....72...76A}, we typically considered a systematic uncertainty of $ \sigma_{\tau_{\mathrm{syst}}} = 1.5\%  \cdot \tau $ on the estimate of the residual background.

The joint likelihood function $ \mathcal{L} $ is a nested product of the binned likelihood function $ \mathcal{L}_{\text{bin},kl} $ (Eq.~\ref{eq:Binned_lkl}) for each dSphs ($N_{\text{dSphs}} = 4$) and their distinct observational datasets $ \bm{\mathcal{D}_{kl}} $ with individual set of IRFs due to different observational conditions or hardware setup of the instrument. It reads as:
\begin{equation}
    \mathcal{L}\left( \langle \sigma v \rangle \right) = \prod_{k=1}^{N_{\text{dSphs}}} \prod_{l=1}^{N_{\text{obs},k}} \Big[ \mathcal{L}_{\text{bin},kl} \left( \langle \sigma v \rangle, \bm{\nu_{kl}} \mid \bm{\mathcal{D}_{kl}} \right) \Big] \times \mathcal{J}_{k} \left( J_{k} \mid J_{\text{o},k}, \sigma_{\log_{10}J_{k}} \right)
\end{equation}
\noindent
where $ N_{\text{obs},k} $ is the number of observations for the $ k $-th dSph and $ \bm{\nu_{kl}} $ represents the set of nuisance parameters different from the \textit{J}-factor affecting the analysis of dataset $ \bm{\mathcal{D}_{kl}} $. Given the importance of the \textit{J}-factors and their uncertainties (see Sec.~\ref{sec:Observations}), we treat the \textit{J}-factors as nuisance parameters using the likelihood $ \mathcal{J}_{k} $ for the \textit{J}-factor of the $ k $-th dSph (ignoring index $ k $ in the following for the sake of clarity)
\begin{equation}
    \mathcal{J} \left( J \mid J_{\text{o}}, \sigma_{\log_{10}J} \right) = \frac{1}{\ln{(10)} J_{\text{o}} \sqrt{2\pi} \sigma_{\log_{10}J} } \exp{\left(-\frac{\left( \log_{10} J - \log_{10} J_{\text{o}} \right)^{2} }{2\sigma_{\log_{10}J}^{2}}\right)},
    \label{eq:J_lkl}
\end{equation}
\noindent
where $ J $ is the true value of the \textit{J}-factor and $ J_{\text{o}} $ is the observed \textit{J}-factor with error $ \sigma_{\log_{10}J} $~\cite{2015PhRvL.115w1301A}. 

In the absence of a branon DM signal, ULs on $\langle \sigma v \rangle$ for all datasets $ \bm{\mathcal{D}} $ are set using a test statistic defined as
\begin{equation}
    \label{Eq:test_statistic}
    \mathrm{TS} = -2 \ln{ \left(\frac{\mathcal{L} \left( \langle \sigma v \rangle ; \widehat{\widehat{\bm{\nu}}} \mid \bm{\mathcal{D}} \right)}{\mathcal{L} \left( \widehat{\langle \sigma v \rangle} ;  \widehat{\bm{\nu}} \mid \bm{\mathcal{D}} \right)}\right)},
\end{equation}
where $\widehat{\langle \sigma v \rangle}$ and $\widehat{\bm{\nu}}$ are the values that globally maximize $\mathcal{L}$, and $\widehat{\widehat{\bm{\nu}}}$ is the set of values that maximize $\mathcal{L}$ for a particular value of $\langle \sigma v \rangle$. In particular, the ULs on $\langle \sigma v \rangle$ are computed by solving $\mathrm{TS}= 2.71 $, where $2.71$ corresponds to a one-sided 95\% confidence level~\cite{ROLKE2005493}. No additional boosts from the presence of substructures~\cite{2007PhRvD..75h3526S} or quantum effects~\cite{2004PhRvL..92c1303H} entered the computation of the final results.

\section{Results}

\label{sec:Results}

\begin{figure}[h]
    \centering
    \begin{subfigure}{.675\textwidth}
        \includegraphics[width=\textwidth]{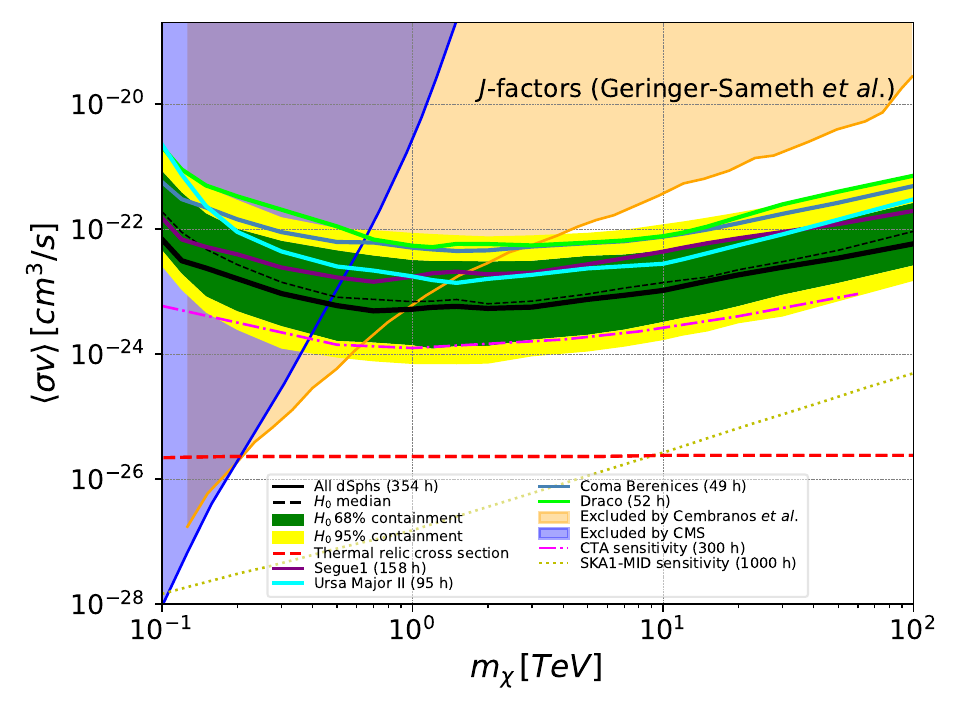}
    \end{subfigure}
    \begin{subfigure}{.675\textwidth}
        \includegraphics[width=\textwidth]{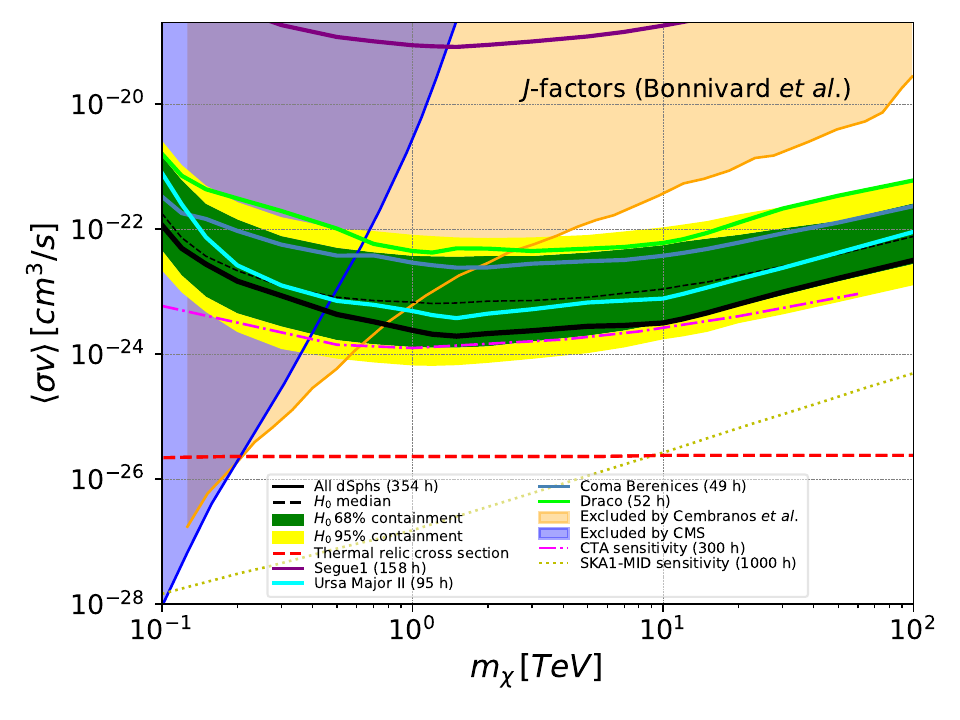}
    \end{subfigure}
\caption{$ 95 \% $ ULs on $ \langle \sigma v \rangle $ for branon DM annihilation from the combined analysis (solid black line) and the analysis of the individual dSphs (Segue~1 purple, Ursa~Major~II cyan, Draco lime green, and Coma~Berenices steel blue). The median and the two-sided $ 68 \% $ and $ 95 \% $ containment bands of the combined analysis are depicted by the dotted black line, green and yellow bands, respectively. The red dashed line indicates the thermal relic cross-section from~\cite{2012PhRvD..86b3506S}. The blue exclusion region represents the tightest constraints to branons model by colliders obtained from CMS data~\cite{2014arXiv1410.8812C} and the orange exclusion region was obtained by Cembranos \emph{et al.} from an analysis~\cite{2017arXiv170909819C} of AMS-02 $ e^{+}e^{-} $ data~\cite{PhysRevLett.110.141102}. They were translated to the $ \langle \sigma v \rangle $ parameter space from~\cite{2020PDU....2700448C}. The estimated branon sensitivity for 300 h observation of Draco with the future CTA is depicted by the purple dashed-dotted line~\cite{2020JCAP...10..041A}. The yellow dotted line represents the estimated sensitivity for 1000 h observation of Draco with the planned SKA assuming the $ W^{+}W^{-} $ annihilation mode~\cite{2020PDU....2700448C}.}
\label{fig:BranonLimits}
\end{figure}

Our likelihood analysis is coherent with all previously reported results~\cite{2011JCAP...06..035A,2012PhRvD..85f2001A,2014JCAP...02..008A,2016JCAP...02..039M} in that no gamma-ray signal has been detected (see also Tab.~\ref{Tab:MAGICObs}). Thus, we present the $ 95 \% $ confidence level upper limits (ULs) on the thermally-averaged cross-section $ \langle \sigma v \rangle $ for branon DM annihilation in a particle mass range from $ \unit[100]{GeV} $ to $ \unit[100]{TeV} $. The ULs are obtained with the before-mentioned combined analysis of multiple dSph observations with the MAGIC telescopes for two different sets of \textit{J}-factors (see Fig. \ref{fig:BranonLimits}). We include systematic uncertainties in the residual background intensity and statistical uncertainties in the \textit{J}-factor in our likelihood analysis. Our strongest limit is $ \langle \sigma v \rangle \simeq \unit[1.9 \times 10^{-24}]{cm^{3}s^{-1}} $ for a $ \sim \unit[1.5]{TeV} $ mass branon DM particle.

We obtain the two-sided $ 68 \% $ and $ 95 \% $ containment bands as well as the median from the distribution of ULs performing the same analysis of 300 fast simulations of the source and background regions assuming no DM signal ($ \langle \sigma v \rangle = 0 $). As expected, our constraints are located within the  $ 68 \% $ containment band for both sets of \textit{J}-factors. The ULs from each individual dSph observation are also depicted in Fig.~\ref{fig:BranonLimits}. For GS15, the combined analysis is dominated by Ursa~Major~II and Segue~1, while the latter target does not have any substantial contribution in the combination for B16. Given the rather huge negative fluctuation of the Ursa~Major~II observation ($-2.1 \sigma$ for a conventional IACT analysis; see Tab.~\ref{Tab:MAGICObs}) and Ursa~Major~II being the most dominant dSph in the analysis for B16, the combined limit for B16 locates significantly below the median. The estimated branon sensitivity for 300~h observation on the dSph Draco with the future Cherenkov Telescope Array (CTA)~\cite{2020JCAP...10..041A} lays an order of magnitude below the median of both presented analyses, as expected.

\begin{figure}[h]
    \centering
    \includegraphics[scale=0.85]{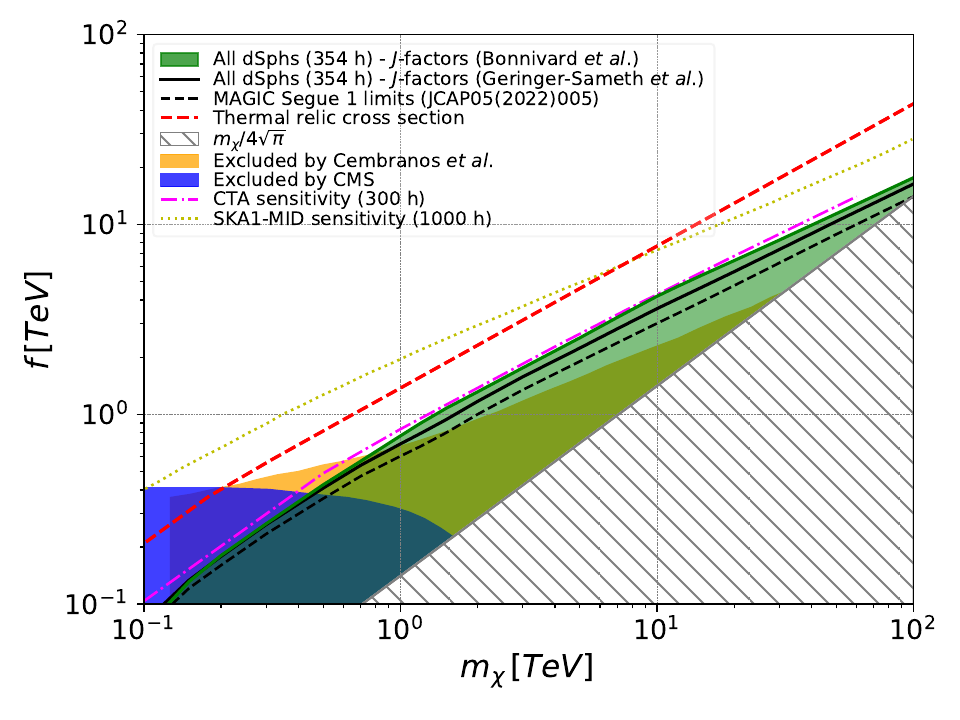}
    \caption{$ 95 \% $ ULs on $ f $ for branon DM annihilation from the combined analysis with the B16 \textit{J}-factors set (green exclusion region) and the GS15 \textit{J}-factors set (solid black line). The ULs previously obtained with the MAGIC observation of Segue~1 are depicted by the dashed black line~\cite{Miener:2022fon,2022arXiv220507055M}. The
grey dashed region depicts the model validity limit in the $ f \left( m_{\chi} \right)$ parameter space.}% The color code for the other instruments has been adopted from Fig.~\ref{fig:BranonLimits}.}
    \label{fig:Branetension_limits}
\end{figure}

We set constraints to the specific parameter space of the branon DM model (see Fig. \ref{fig:Branetension_limits}), i.e.~the tension of the brane  $f(m_\chi)$ versus the branon DM mass ranging from $ \unit[100]{GeV} $ to $ \unit[100]{TeV} $, by translating our $ \langle \sigma v \rangle $ ULs to constraints on $ f $~\cite{Miener:2022fon,2022arXiv220507055M}. The combined analysis of this work allows us to exclude a significantly larger portion of the brane tension parameter space than previous branon ULs in the literature by CMS~\cite{2014arXiv1410.8812C}, Cembranos \emph{et al.}~\cite{2017arXiv170909819C} with AMS-02 $ e^{+}e^{-} $ data~\cite{PhysRevLett.110.141102}, and MAGIC with the observation of Segue~1~\cite{Miener:2022fon,2022arXiv220507055M}, for both sets of \textit{J}-factors, GS15 and B16. Although, the constraints obtained by Cembranos \emph{et al.}~\cite{2017arXiv170909819C} with AMS-02 data~\cite{PhysRevLett.110.141102} would require an updated analysis including more recent AMS-02 data from~\cite{AGUILAR20211}.

\section{Discussion and conclusions}

\label{sec:DisCon}

\begin{figure}[h]
    \centering
    \includegraphics[scale=0.75]{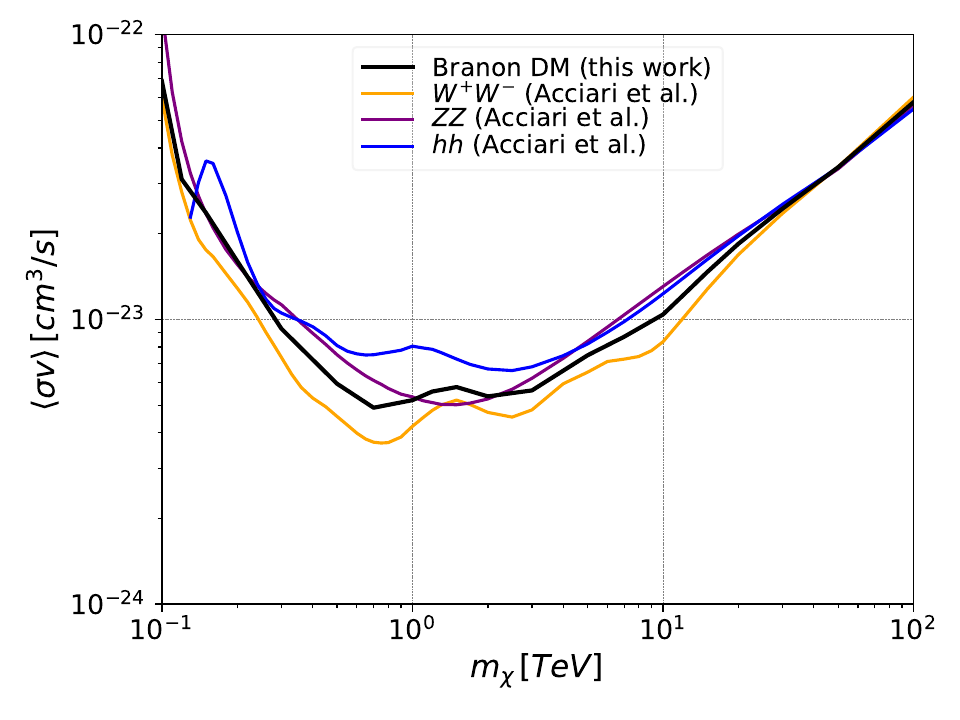}
    \caption{Comparison of the $ 95 \% $ ULs on $ \langle \sigma v \rangle $ for branon DM annihilation (black) and the dominant annihilation mode $ W^{+}W^{-} $ (yellow), $ ZZ $ (purple), and $ hh $ (blue) in the branon DM model at VHE taken from~\cite{MAGIC:2021mog}.}
    \label{fig:Branon_limits_withMAGIC}
\end{figure}

The branon DM exclusion limits are compared with the dominant annihilation mode $ W^{+}W^{-} $, $ ZZ $, and $ hh $ in the branon DM model at VHE. The model-independent ULs in Fig.~\ref{fig:Branon_limits_withMAGIC} are taken from~\cite{MAGIC:2021mog}, which rely on the same dSph datasets with the same analysis scheme. In~\cite{MAGIC:2021mog}, the \textit{J}-factor values of GS15 were used to compute the model generic DM exclusion limits. The ULs for branon DM annihilation are enclosed by the dominant annihilation channels verifying the correctness of the performed analysis (see Fig.~\ref{fig:Branon_limits_withMAGIC}).

In comparison with the model-independent search for DM~\cite{MAGIC:2021mog}, this work is not only capable of constraining the thermally-averaged cross-section $ \langle \sigma v \rangle $, but also the brane tension $ f(m_\chi) $. The tension of the brane can be affected by various factors, such as the curvature of the spacetime, the number of dimensions in the brane, and the type of matter present in the brane. It plays a crucial role in theories such as string theory and brane cosmology, which aim to explain the fundamental nature of our universe. It helps to determine the dynamics and behavior of branes within a given spacetime.

This work supersedes the exclusion limits from CMS~\cite{2014arXiv1410.8812C} and Cembranos \emph{et al.}~\cite{2017arXiv170909819C} with AMS-02 data even at the upper edge of the sub-TeV DM mass range leading to the most constraining branon DM limits above $ \sim \unit[700]{GeV}$ by combining all major dSph observations of the MAGIC telescopes. Even more stringent and robust exclusion limits of the branon DM annihilation over a wider range of branon DM masses can be achieved in the framework of multi-instrument and multi-messenger DM searches~\cite{Oakes:2019, Armand:2021} by performing a global branon DM search with a joint analysis of observational data from different ground/space-based gamma-ray and neutrino telescopes. Future instruments, such as CTA and SKA, will probe even a larger fraction of the exclusion region, providing valuable complementary information in both gamma-ray and radio observations, respectively.

\acknowledgments
T. Miener: Principal investigator, MAGIC data analysis, publication coordination; D. Kerszberg: MAGIC data analysis, publication coordination; V. Gammaldi: Branon Dark Matter theory, interpretation J. Rico: Statistical analysis supervision and software framework development; D. Nieto: supervision and coordination, interpretation. The rest of the authors have contributed in one or several of the following ways: design, construction, maintenance, and operation of the instrument(s); preparation and/or evaluation of the observation proposals; data acquisition, processing, calibration and/or reduction; production of analysis tools and/or related Monte Carlo simulations; discussion and approval of the contents of the draft.
We would like to thank the Instituto de Astrof\'{\i}sica de Canarias for the excellent working conditions at the Observatorio del Roque de los Muchachos in La Palma. The financial support of the German BMBF, MPG and HGF; the Italian INFN and INAF; the Swiss National Fund SNF; the grants PID2019-104114RB-C31, PID2019-104114RB-C32, PID2019-104114RB-C33, PID2019-105510GB-C31, PID2019-107847RB-C41, PID2019-107847RB-C42, PID2019-107847RB-C44, PID2019-107988GB-C22, PID2022-136828NB-C41, PID2022-137810NB-C22, PID2022-138172NB-C41, PID2022-138172NB-C42, PID2022-138172NB-C43, PID2022-139117NB-C41, PID2022-139117NB-C42, PID2022-139117NB-C43, PID2022-139117NB-C44 funded by the Spanish MCIN/AEI/ 10.13039/501100011033 and “ERDF A way of making Europe”; the Indian Department of Atomic Energy; the Japanese ICRR, the University of Tokyo, JSPS, and MEXT; the Bulgarian Ministry of Education and Science, National RI Roadmap Project DO1-400/18.12.2020 and the Academy of Finland grant nr. 320045 is gratefully acknowledged. This work was also been supported by Centros de Excelencia ``Severo Ochoa'' y Unidades ``Mar\'{\i}a de Maeztu'' program of the Spanish MCIN/AEI/ 10.13039/501100011033 (CEX2019-000920-S, CEX2019-000918-M, CEX2021-001131-S) and by the CERCA institution and grants 2021SGR00426 and 2021SGR00773 of the Generalitat de Catalunya; by the Croatian Science Foundation (HrZZ) Project IP-2022-10-4595 and the University of Rijeka Project uniri-prirod-18-48; by the Deutsche Forschungsgemeinschaft (SFB1491) and by the Lamarr-Institute for Machine Learning and Artificial Intelligence; by the Polish Ministry Of Education and Science grant No. 2021/WK/08; and by the Brazilian MCTIC, CNPq and FAPERJ. This work was supported by the Grant RYC2021-032552-I funded by MCIN/AEI/10.13039/501100011033 and by the European Union NextGenerationEU/PRTR.\newline
VG’s contribution to this work has been supported by Juan de la Cierva-Incorporaci\'on IJC2019-040315-I grant, and by the PGC2018-095161-B-I00, PID2022-139841NB-I00 and CEX2020-001007-S projects, both funded by MCIN/AEI/10.13039/501100011033 and by "ERDF A way of making Europe". VG  thanks J.A.R. Cembranos for useful discussions.\newline

\bibliographystyle{JHEP}
\bibliography{main}

\end{document}